Correspondence and requests for materials should be addressed to *K.K. (kudo@science.okayama-u.ac.jp) or **M.N. (nohara@science.okayama-u.ac.jp)

# Emergence of superconductivity at 45 K by lanthanum and phosphorus co-doping of $CaFe_2As_2$

Kazutaka Kudo*, Keita Iba, Masaya Takasuga, Yutaka Kitahama, Jun-ichi Matsumura, Masataka Danura, Yoshio Nogami & Minoru Nohara**

Department of Physics, Okayama University, Okayama 700-8530, Japan

**Co-doping of lanthanum and phosphorus in $CaFe_2As_2$ induces superconductivity at 45 K. This superconducting transition temperature is higher than the 38 K transition in $Ba_{1-x}K_xFe_2As_2$, which is the maximum found thus far among the 122 phases. Superconductivity with a substantial shielding volume fraction was observed at $0.12 \leq x \leq 0.18$ and $y = 0.06$ in $Ca_{1-x}La_xFe_2(As_{1-y}P_y)_2$. The superconducting phase of the present system seems to be not adjacent to an antiferromagnetic phase.**

Iron arsenide superconductors seem to occur in close proximity to an antiferromagnetic (AF) phase.[1,2,3] Density functional theory calculations have pointed out the importance of spin fluctuations, resulting from Fermi surface nesting between hole and electron pockets, for the emergence of superconductivity.[4,5] The superconducting transition temperature $T_c$ has exceeded 50 K in doped REFeAsO, where RE corresponds to a rare earth element.[6,7,8]

However, recent reports by Iimura et al.[9] and Sun et al.[10] suggest that the fascinating superconducting phase is also found separated from the AF phase in this class of materials. Iimura et al.[9] reported that two domes appear in $T_c$ versus $x$ for $LaFeAsO_{1-x}H_x$. The first dome exhibits $T_c$ = 26 K in the vicinity of the AF phase, which was recognized as the maximum $T_c$ of the electron-doped $LaFeAsO$.[11] Surprisingly, a much higher $T_c$ = 36 K emerges in the second dome, far from the AF phase. Moreover, Sun et al.[10] reported the re-emergence of high temperature superconductivity in $Tl_{0.6}Rb_{0.4}Fe_{1.67}Se_2$, $K_{0.8}Fe_{1.7}Se_2$, and $K_{0.8}Fe_{1.78}Se_2$ as a function of pressure. At low pressures, superconductivity in those compounds coexists with AF order, as reported by Guo et al.[12] With increasing pressure, $T_c$ decreases from a maximum of 32 K at 1 GPa to zero at 9.8 GPa[10], above which AF order disappears.[12] Upon further increase in pressure, a second superconducting phase with $T_c$ = 48 K suddenly emerges above 11.5 GPa in the paramagnetic phase.[10] These results suggest a greater flexibility in developing iron-based superconductors. A similar superconducting phase that is separated from the AF phase is expected in systems other than $LaFeAsO_{1-x}H_x$[9] and $(Tl,Rb,K)_{1-x}Fe_{2-y}Se_2$.[10]

We focus on the RE-doped $CaFe_2As_2$ (abbreviated to 122) system as a candidate material with a superconducting phase separated from an AF phase. Very recently, in the vicinity of the AF phase, a superconducting phase with $T_c$ = 40-49 K was suggested in RE-doped $CaFe_2As_2$.[13,14,15] Though bulk superconductivity has not yet been confirmed, the results strongly suggest that the 122 phase may exhibit a $T_c$ as high as 50 K. This value is higher than $T_c$ = 38 K in $Ba_{1-x}K_xFe_2As_2$[16] which is the highest value yet reported among the 122-type bulk superconductors. Similar results were almost simultaneously reported by three groups: Saha et al. reported $T_c$ = 47 K in $Ca_{1-x}Pr_xFe_2As_2$;[13] Gao et al. reported $T_c$ = 42.7 K in $Ca_{1-x}La_xFe_2As_2$;[14] and Lv et al. reported $T_c$ = 49 K in $Ca_{1-x}Pr_xFe_2As_2$.[15] These values of $T_c$ were determined from the onset of the resistive transition.

In the reports by Saha et al.[13] and Gao et al.,[14] diamagnetic behavior is not clearly observed in the temperature dependence of the magnetization $M$ around $T_c$. Instead, a visible drop in $M$ occurs below ~20 K, suggesting that a minority phase possesses a high $T_c$. On the other hand, Lv et al.[15] showed a clear diamagnetic signal around 40 K in a magnetic field of 1 Oe, in addition to a subsequent transition at 20 K. However, the signal around 40 K was completely suppressed by the application of a tiny magnetic field of 20 Oe. Lv et al.[15] proposed that interfacial superconductivity with $T_c$ = 49 K occurs at the grain boundary.

In this paper, we report the emergence of superconductivity at 45 K by La and P co-doping of $CaFe_2As_2$. $T_c$ = 45 K is higher than 38 K in $Ba_{1-x}K_xFe_2As_2$.[16] A substantial shielding volume fraction was observed at $0.12 \leq x \leq 0.18$ with $y$ = 0.06 in $Ca_{1-x}La_xFe_2(As_{1-y}P_y)_2$. The superconducting phase of the present system seems to be far from the AF phase in the electronic phase diagram.

**Results**

**Lattice parameters and cell volume.**

In the phosphorus-free $Ca_{1-x}La_xFe_2As_2$, the $a$ parameter gently increases and the $c$ parameter slightly decreases with increasing $x$, as shown in Figure 1. This result is consistent with Saha et al.[13] Upon phosphorus doping of $Ca_{1-x}La_xFe_2As_2$, the $a$ parameter increases and the $c$ parameter decreases, resulting in a reduced cell volume. This decrease in cell volume is expected since the ionic radius of P is smaller than As. Kasahara et al.[17] have shown in $CaFe_2(As_{1-x}P_x)_2$ that the reduction in volume leads to successive phase transitions in which the AF phase, the superconducting phase, and the non-superconducting phase appear, in this order. Thus, isovalent phosphorus doping separates the system from the AF phase. This same effect is also expected in $Ca_{1-x}La_xFe_2(As_{1-y}P_y)_2$.

**Superconductivity.**

In fact, phosphorus-doped $Ca_{1-x}La_xFe_2(As_{0.94}P_{0.06})_2$ does not show an AF transition. As shown in the inset of Figure 2, phosphorus-free $Ca_{1-x}La_xFe_2As_2$ ($x$ = 0.12) exhibits a jump in the temperature dependence of the electrical resistivity, characteristic of an AF transition.[19,20] On the other hand, no jump was observed in phosphorus-doped $Ca_1$- $_x$La$_x$Fe$_2$(As$_{0.94}$P$_{0.06}$)$_2$ ($x$ = 0.12). Instead, the electrical resistivity exhibits superconductivity with $T_c$ as high as 40 K. A maximum $T_c$ of 45 K was observed around $x$ = 0.15 – 0.18 with $y$ = 0.06. Figure 2 shows the temperature dependence of the electrical resistivity for Ca$_{1-x}$La$_x$Fe$_2$(As$_{0.94}$P$_{0.06}$)$_2$ ($x$ = 0.17). The resistivity decreases with decreasing temperature, starts to drop sharply at 48 K, and becomes negligibly small below 45 K. This result suggests that the sample becomes superconducting below 45 K.

Superconductivity at 45 K was also evident from the temperature dependence of the magnetization, shown in Figure 3. Ca$_{1-x}$La$_x$Fe$_2$(As$_{0.94}$P$_{0.06}$)$_2$ ($x$ = 0.17) shows clear diamagnetic behavior below 45 K. The estimated shielding volume fraction, VF, is as large as 23% at 20 K and 75% at 5 K. Contrary to previous reports,[13,14,15] there is no second transition at 20 K, and the VF is robust against any increase in magnetic field, as shown in the inset of Figure 3. These results do not entirely ensure the occurrence of bulk superconductivity at 45 K because the transition is broad and the magnetization continues to decrease with decreasing temperature. Nonetheless, the present data strongly suggest the existence of a bulk superconducting phase with $T_c$ = 45 K in Ca$_{1-x}$La$_x$Fe$_2$(As$_{0.94}$P$_{0.06}$)$_2$.

**Phase diagram.**

Figure 4a shows the $T$-$x$-$y$ phase diagram of Ca$_{1-x}$La$_x$Fe$_2$(As$_{1-y}$P$_y$)$_2$ for $y$ = 0.00, 0.03, and 0.06, derived from electrical resistivity and magnetization measurements. The AF phase is almost suppressed by 3% phosphorus doping. It is found that the superconducting phase (colored in blue) seems to be not adjacent to the AF phase. Superconductivity at 40-45 K was observed at 0.12 ≤ $x$ ≤ 0.18 with $y$ = 0.06 in Ca$_{1-x}$La$_x$Fe$_2$(As$_{1-y}$P$_y$)$_2$. The substantial VF in these samples can be seen in Figure 4b. On the other hand, near the AF phase, phosphorus-free Ca$_{1-x}$La$_x$Fe$_2$As$_2$ and Ca$_{1-x}$La$_x$Fe$_2$(As$_{0.97}$P$_{0.03}$)$_2$ with less phosphorus exhibit traces of superconductivity, with a zero resistivity accompanied by a quite small VF.

**Discussion**

At $x < 0.12$, $Ca_{1-x}La_xFe_2(As_{0.94}P_{0.06})_2$ exhibits trace superconductivity. A superconducting phase with a large VF suddenly emerges at $x \geq 0.12$ and is narrow in width ($\Delta x = 0.06$). Here, the question of the origin for the sudden cut-off of the superconducting phase arises. The lattice collapse transition in $CaFe_2As_2$[17,18,21,22] might provide a clue. $CaFe_2As_2$ undergoes a lattice collapse transition from an uncollapsed tetragonal (ucT) to a collapsed tetragonal (cT) phase along with a 10% decrease in $c$ with the application of pressure or chemical doping. It is well known that superconductivity abruptly disappears with the appearance of the ucT-cT transition.[17,18,21,22] The sudden cut-off of the superconducting phase at $x < 0.12$ might be explained in terms of the ucT-cT transition. Structural studies are thus necessary in the present system.

Another question that arises is on the origin of the emergence of high $T_c$ superconductivity separate from the AF phase. The second dome in $T_c$ in the $T$-$x$ phase diagram of $LaFeAsO_{1-x}H_x$ may provide a hint.[9] Based on density functional theory calculations, Iimura et al.[9] have shown that the three Fe 3d bands ($xy$, $yz$, and $zx$) become degenerate when $T_c$ exhibits a maximum in the second dome, whereas Fermi surface nesting is weakened there. They have implied that band degeneracy is a key ingredient to induce high $T_c$ in the second dome, while spin fluctuations are important in the first dome. A similar scenario may be active in $Ca_{1-x}La_xFe_2(As_{1-y}P_y)_2$. Further angle-resolved photoemission spectroscopy measurements and band structure calculations are expected to be useful for understanding superconductivity in the present system.

So far, systematic chemical substitutions have been conducted in the 122 type $AEFe_2As_2$ (AE = alkaline earth elements). It has been found that bulk superconductivity appears upon substituting alkali metals for AE (hole doping),[16] the RE element La for AE (= Sr) (electron doping),[23] transition metal elements such as Co[24,25] and Ni[26,27] for Fe (electron doping), and P for As (isovalent doping).[17,28,29,30,31]

The highest resulting $T_c$ is 38 K in $Ba_{1-x}K_xFe_2As_2$.[16] There seems to be no further variety of chemical doping in the 122 phase. However, there is still some room left for improvement in terms of co-doping. Sole P doping of the As site shrinks the cell volume, resulting in bulk superconductivity at 15 K in $CaFe_2(As_{1-x}P_x)_2$.[17] Sole La doping of the Ca site induces electron carriers, leading to trace superconductivity at 40 K in $Ca_{1-x}La_xFe_2As_2$.[13,14] In this study, La and P co-doped $Ca_{1-x}La_xFe_2(As_{1-y}P_y)_2$ exhibits $T_c$ = 45 K, which is higher than 38 K. It is considered that the simultaneous tuning of the concentration of electron carriers and the reduction in cell volume optimizes superconductivity. Our result suggests that the co-doping technique will lead to higher superconducting transition temperatures in the iron-based pnictides.

In conclusion, we demonstrate the emergence of superconductivity at 45 K by La and P co-doping of $CaFe_2As_2$. $T_c$ = 45 K is higher than the value of 38 K found in $Ba_{1-x}K_xFe_2As_2$. The magnetization shows a substantial shielding volume fraction at $0.12 \leq x \leq 0.18$ with $y$ = 0.06 in $Ca_{1-x}La_xFe_2(As_{1-y}P_y)_2$. The superconducting phase in the present system looks to be separated from the antiferromagnetic phase.

**Methods**

   **Preparation and characterization of samples.**

Single crystals of $Ca_{1-x}La_xFe_2(As_{1-y}P_y)_2$ were grown using a self-flux method.[18] A mixture with a ratio of Ca:La:FeAs:Fe:P = 1-$x$:$x$:4-4$y$:4$y$:4$y$ was placed in an alumina crucible, sealed in an evacuated quartz tube, slowly heated to 1100°C, and cooled to 1050°C at a rate of 1.25°C/h followed by furnace cooling. Single crystals with a typical dimension of 1.5 x 1.5 x 0.05 mm$^3$ were mechanically isolated from the flux. The results of powder X-ray diffraction at room temperature, performed using a Rigaku RINT-TTR III X-ray diffractometer with Cu K$\alpha$ radiation, showed that all specimens are single phase and possess the $CaFe_2As_2$ structure. Energy dispersive X-ray spectrometry (EDS) was used to determine the lanthanum content $x$ and the phosphorus content $y$. The lattice parameters $a$ and $c$ were estimated at room temperature using a Rigaku Single Crystal X-ray Structural Analyzer (Varimax with Saturn).

**Electrical resistivity and magnetization measurements.**

The electrical resistivity $\rho_{ab}$ (parallel to the *ab*-plane) measurements were carried out by a standard DC four-terminal method in a Quantum Design PPMS. The magnetization *M* was measured using a SQUID magnetometer (Quantum Design MPMS SQUID-VSM).

**Acknowledgments**

Part of this work was performed at the Advanced Science Research Center, Okayama University. This work was partially supported by a Grant-in-Aid for Young Scientists (B) (23740274) from the Japan Society for the Promotion of Science (JSPS) and the Funding Program for World-Leading Innovative R&D on Science and Technology (FIRST Program) from JSPS.

**Author contributions**

K.K. and M.N. conceived and planed the research. K.I., Y.K, J.M., and K.K. synthesized single crystals. K.I., M.T., M.D., and K.K. carried out electrical resistivity and magnetization measurements. K.K., K.I., M.T., and Y.N. characterized single crystals using X-ray diffraction. K.K. and M.N. discussed the results and wrote the manuscript.

**Additional information**

Competing financial interests: The authors declare no competing financial interests.

**Figure caption**

**Figure 1 Lattice parameters and cell volume of $Ca_{1-x}La_xFe_2(As_{1-y}P_y)_2$.** $x$ dependence of (**a**) *a* parameter, (**b**) *c* parameter, and (**c**) cell volume for $y = 0$ and 0.06. The vertical error bars signify the standard deviation of the experimental data. The data for the sample with $x = y = 0$ are quoted from Refs. 17 and 18, and those with $x = 0$ and $y = 0.06$ from Ref. 17.

**Figure 2 Temperature dependence of electrical resistivity $\rho_{ab}$ for $Ca_{1-x}La_xFe_2(As_{1-y}P_y)_2$ ($x = 0.17$ and $y = 0.06$).** The inset shows that for $x = 0.12$ with $y = 0.00$ and 0.06. The arrow indicates the antiferromagnetic/tetragonal-orthorhombic structural transition.

**Figure 3 Temperature dependence of magnetization *M* divided by magnetic field *H*, *M/H* for $Ca_{1-x}La_xFe_2(As_{1-y}P_y)_2$ ($x = 0.17$ and $y = 0.06$).** The inset shows *M/H* vs *T* for several magnetic fields up to 60 Oe for samples within the same batch.

**Figure 4. Phase diagram and shielding volume fraction for $Ca_{1-x}La_xFe_2(As_{1-y}P_y)_2$.** (**a**) *T-x-y* electronic phase diagram for $y = 0.00$, 0.03, and 0.06. $T_c(M/H)$ is the bulk superconducting transition temperature determined from the temperature dependence of magnetization. $T_c(\rho_{zero})$ is the temperature below which the electrical resistivity becomes negligibly small. $T_N$ is the antiferromagnetic/structural transition temperature, determined as the temperatures below which the electrical resistivity exhibits a jump.[19,20] O and T indicate orthorhombic and tetragonal phases, respectively. AF and SC indicate the antiferromagnetic and the superconducting phases, respectively. SC trace indicates the non-bulk superconducting phase. (**b**) *x* dependence of shielding volume fraction VF at $T = 5$ and 20 K for $y = 0.00$, 0.03, and 0.06.

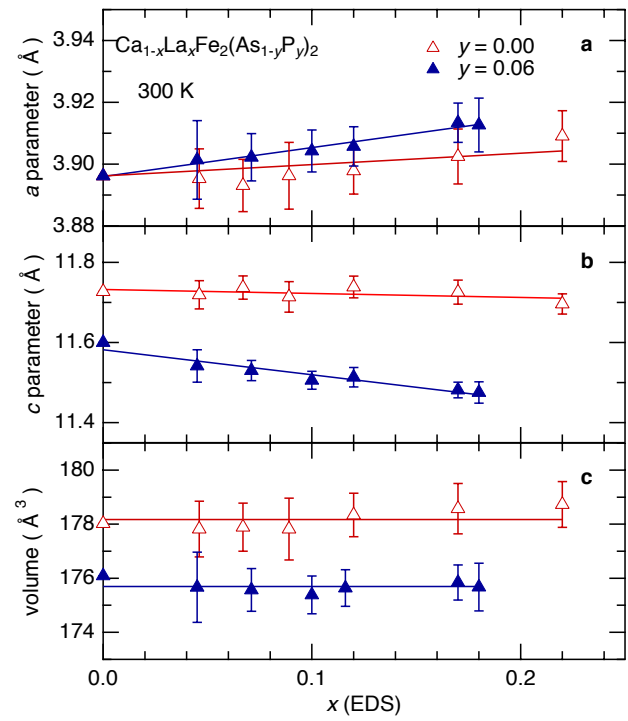

Figure 1

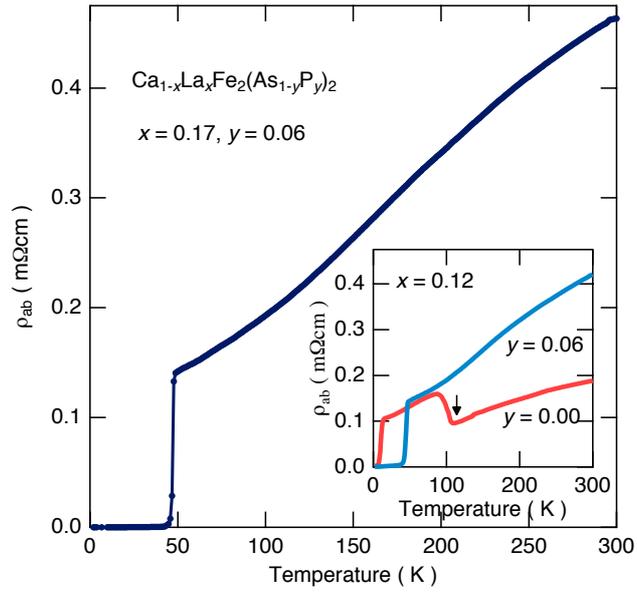

Figure 2

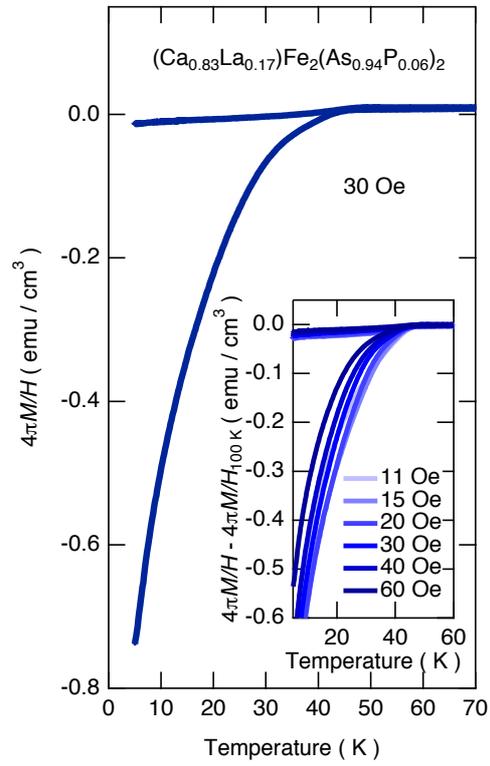

Figure 3

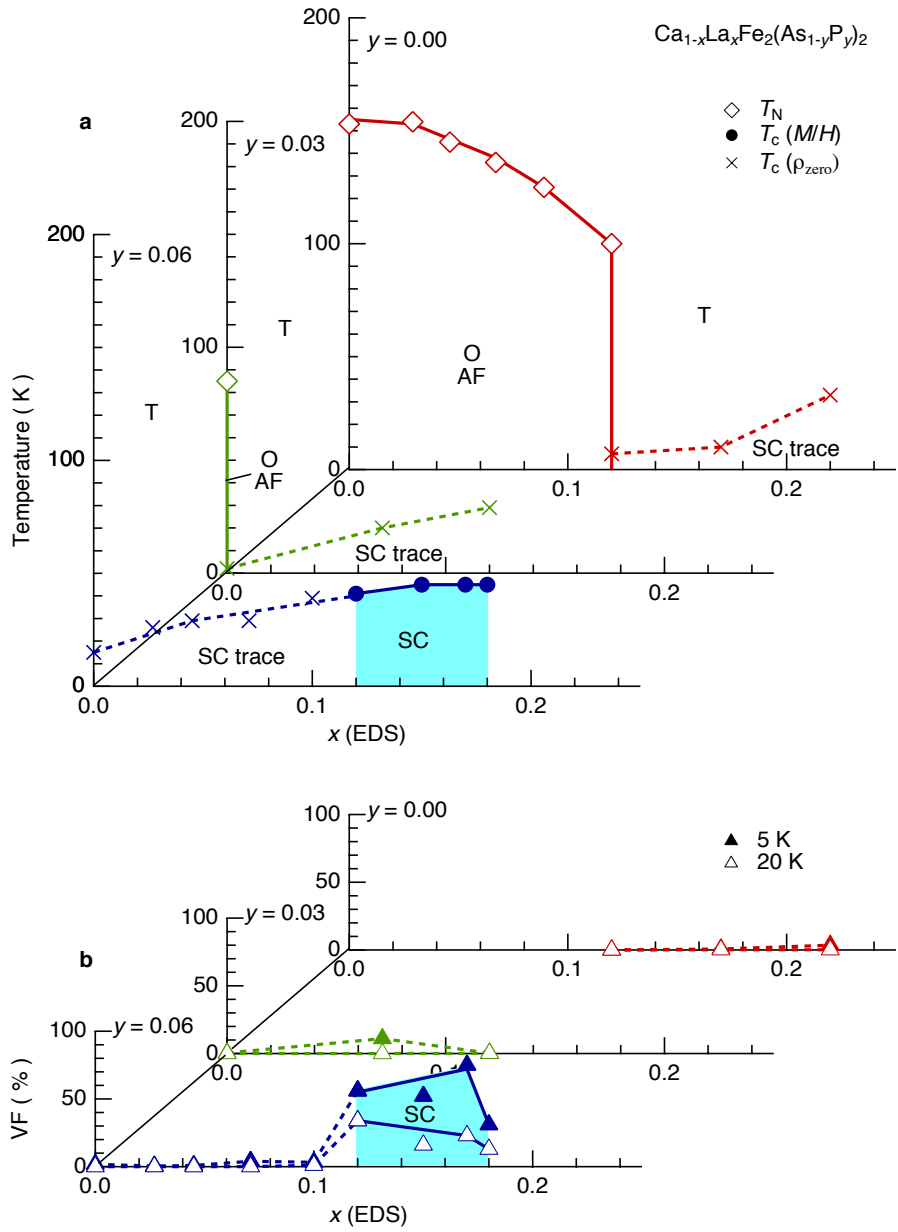

Figure 4